\documentclass{article} 
\usepackage{style/arxiv}

\usepackage{url}
\usepackage{import}
\usepackage{multirow}
\usepackage{subcaption}
\usepackage{graphicx}
\usepackage{amsmath}

\begin{document}

\title{Evaluating Eye Tracking Signal Quality with Real-time Gaze Interaction Simulation}

\author{ 
{\hspace{1mm}Mehedi Hasan Raju}\thanks{corresponding author} \\
Texas State University\\
San Marcos, Texas, 78640, USA\\
\texttt{m.raju@txstate.edu} \\
\And
{\hspace{1mm}Samantha~Aziz} \\
Texas State University\\
San Marcos, Texas, 78640, USA\\
\texttt{sda69@txstate.edu} \\
\And
{\hspace{1mm}Michael J. Proulx} \\
University of Bath\\
Bath BA2 7AY, UK \\
\texttt{m.j.proulx@bath.ac.uk} \\
\And
{\hspace{1mm}Oleg V. Komogortsev} \\
Texas State University\\
San Marcos, Texas, 78640, USA\\
\texttt{ok11@txstate.edu} \\
}
\date{}

\maketitle
\begin{abstract}



We present a real-time gaze-based interaction simulation methodology using an offline dataset to evaluate the eye-tracking signal quality.
This study employs three fundamental eye-movement classification algorithms to identify physiological fixations from the eye-tracking data. 
We introduce the Rank-1 fixation selection approach to identify the most stable fixation period nearest to a target, referred to as the trigger-event.
Our evaluation explores how varying constraints impact the definition of trigger-events and evaluates the eye-tracking signal quality of defined trigger-events.
Results show that while the dispersion threshold-based algorithm identifies trigger-events more accurately, the Kalman filter-based classification algorithm performs better in eye-tracking signal quality, as demonstrated through a user-centric quality assessment using user- and error-percentile tiers.
Despite median user-level performance showing minor differences across algorithms, significant variability in signal quality across participants highlights the importance of algorithm selection to ensure system reliability.

\end{abstract}

\keywords {Eye Movement, Gaze-interaction, Simulation, Classification Algorithms, Real-time}

\section{INTRODUCTION}

Recent innovations in consumer electronics, such as Meta's Orion  \cite{MetaOrion} and Apple's Vision Pro  \cite{AppleVision}, indicate a shift towards the mainstream adoption of eye-tracking in consumer electronics.
These devices use eye-tracking to enable intuitive gaze-based controls, optimize rendering with foveated techniques, and create more immersive, efficient experiences. 
For instance, Apple introduced eye-tracking as an accessibility feature in iOS 18  \cite{ios18}, enabling users to interact with their devices through gaze, i.e., gaze-based interaction,
thus making gaze interaction available to billions of iOS users, something that was never done before.

Gaze-based interaction utilizes eye movements to control devices and systems, transforming the user's gaze into an intuitive input modality  \cite{stampe1995selection}.
Prior research has highlighted the potential of gaze-based interaction to enhance the efficiency of human-computer interaction (HCI), particularly as a medium for target selection, due to its natural and direct nature \cite{majaranta_2014, Actigaze, ware1986evaluation, kumar2007eyepoint}.
Recent research validates this claim as gaze-based interaction is employed across various devices, including computers \cite{majaranta2002twenty}, smartphones  \cite{holland2012eye}, virtual reality (VR) headsets  \cite{fernandes2023leveling, piumsomboon2017exploring}, and augmented reality (AR) systems  \cite{bektacs2024gaze, shi2023exploring}.
As an input method, gaze interaction offers both benefits and limitations. 
Since gaze inherently gathers visual information, gaze location can effectively indicate the user's focus of attention  \cite{just1976eye}. 
One of its primary advantages is its naturalness. 
Users can easily focus on the target simply by looking at them, faster and less physically demanding, making gaze an intuitive means of interaction  \cite{jacob1995eye, stampe1995selection}.

Despite the many benefits of gaze-based interaction, several challenges remain, particularly regarding the accuracy of gaze detection. 
Accurate gaze detection refers to the system’s ability to pinpoint the exact location on the screen where the user is gazing, such as a specific object or target. 
The better the accuracy, the more effectively the system can respond to the user’s gaze. 
Conversely, poor accuracy can lead to incorrect target selection, disrupting the user experience by causing slower, less efficient interactions and increasing user frustration. 
Furthermore, analyzing eye movements during real-time interactions with dynamic stimuli presents additional challenges such as the noisier nature of the real-time gaze data, and synchronization issues between real-time eye movements and events  \cite{Duchowski_book}. 

To address these challenges, we simulate real-time gaze interaction using the offline GazeBase dataset.
This dataset provides standardized eye movement data with high spatial accuracy, collected in a controlled environment.
The part of the dataset employed in our study includes both gaze data and relative target data, allowing us to precisely control and align gaze-target interactions, mitigating synchronization issues and enabling more consistent and reliable measurements. 
This approach ensures consistency, replicability, and valid comparisons across studies eliminating the inherent variability and delays encountered in real-time scenarios.
Most importantly, the GazeBase dataset allows us to analyze data from 322 subjects, a scale that is often difficult to achieve with interaction-based datasets, which typically involve a much smaller number of participants. 

Motivated by these considerations, we develop a novel methodology for simulating real-time gaze data from an offline dataset to simulate real-time gaze interaction.
This approach introduces the Rank-1 Fixation Selection method, which interprets fixations as a user's intent to interact with the system through gaze.
We adapted real-time versions of three foundational eye movement classification algorithms to reliably identify physiological fixations from eye-tracking data as these fixations are the core unit around which the interactions are scheduled.
Our methodology further evaluates the effectiveness of these simulated interactions. 
Additionally, we assess the eye-tracking signal quality for interaction using a user-centric quality evaluation, employing user- and error-percentile tiers to evaluate performance in the gaze-based system.
Finally, our evaluations incorporate constraints to ensure robustness and applicability across different settings and use cases in gaze-based systems.

\section{PRIOR WORK}

\subsection{Prior Work on Eye Movement Classification Algorithms}

Threshold-based eye movement classification algorithms have been fundamental in developing a simple yet effective approach to categorizing different eye movement types such as fixation and saccade.
The initial threshold-based algorithms, such as Velocity Threshold-based Identification (IVT) and Dispersion Threshold-based Identification (IDT) were developed and provided a foundational approach for automating eye-tracking data classification  \cite{Salvucci, Duchowski_book}.
For instance, a self-adaptive version of the IVT algorithm was developed to detect microsaccades better  \cite{Adaptive_ivt}, while Nyström et al.  \cite{Nystrom_adaptive} introduced an adaptive algorithm to accurately classify fixations, saccades, and glissades by dynamically adjusting classification thresholds to account for differences in viewers, recording conditions, and varying levels of noise. 
Similarly, Ralf et al. proposed an automated velocity-based algorithm that used minimum determinant covariance estimators and control chart procedures to identify fixations  \cite{Van_ibt}.

Over time, more advanced threshold-based methods were developed to handle noisier data and facilitate real-time analysis.
For example, Sauter et al. proposed an algorithm to classify saccades from other types of eye movements using the Kalman filter coupled with $\chi^2$  \cite{sauter1991analysis}.
Komogortsev extended this approach in  \cite{Komorgortsev_kalman}, incorporating velocity and temporal thresholds to identify fixation from smooth pursuits.
This approach with an integrated chi-squared threshold significantly improved accuracy over the traditional IVT method as demonstrated by Koh et al.  \cite{Koh2009, Koh_real}.
The Kalman Filter-based Identification (IKF) method further enhanced the ability to process noisier datasets and enabled real-time analysis  \cite{Komorgortsev_kalman, komogortsev_kalman_chi}.

In addition to threshold-based algorithms, a wide array of eye-movement classification methods has emerged in the literature, employing approaches such as variance and covariance analysis  \cite{Veneri_automatic}, probabilistic models like Hidden Markov Models  \cite{Salvucci, Kasneci2014}, Bayesian theory  \cite{Santini_bayesian, bayesian}, and machine learning techniques  \cite{zemblys2018using, Larsson, Elmadjian_omc, Fuhl_rule}.
More detailed reviews and evaluations about prior classification algorithms can be found in  \cite{andersson2017one, survey}

Despite the many available options, our study will focus on three algorithms--- IVT, IDT, and IKF, known for their simplicity, fundamental role in eye movement classification, and applicability of real-time use  \cite{Duchowski_book}.
These algorithms operate on relatively simple mathematical models (thresholds for IVT and IDT and state estimation for IKF), which minimizes their computational complexity.
Moreover, these algorithms have been foundational in eye movement research for decades, meaning there is a wealth of empirical research to guide their use and modification  \cite{andersson2017one, Nystrom_adaptive, olsen2012tobii, olsen2012tobii, toivanen2016advanced, zhang2010new}.
Our investigation will evaluate the performance of these algorithms in real-time scenarios under various conditions, offering a comparative analysis of their strengths and limitations in specific use cases. 
It will provide a comparative analysis highlighting their strengths and limitations in specific scenarios.
These methods are favored for their simplicity, low computational requirements, and feasibility to be applied in modern real-time applications.

\subsection{Prior Work on Gaze-based Interaction}

Gaze-based interaction is a promising method for HCI because it is natural and non-intrusive, allowing users to control systems with minimal physical effort  \cite{zhang2019evaluation, vidal_natural, dondi2023gaze}. 
Jacob  \cite{jacob1991use} demonstrated early on that gaze could be an effective input method, faster and less physically demanding than traditional devices like mouse or keyboard. 
Since then, gaze-based interaction has found applications in many areas, including assistive technologies for people with physical impairments  \cite{impairments} and interactive gaming  \cite{gaming}.
With advances in eye-tracking technology, gaze interaction has become more accurate and responsive, making it suitable for mainstream use, such as in smartphones  \cite{holland2012eye, ios18} and extended reality systems  \cite{holmqvist2011eye, khamis2018vrpursuits}.

One key challenge in gaze-based interaction is avoiding unintentional selection when users look around commonly referred to as the Midas touch  \cite{Midas_touch1, Midas_touch2}.
The most common method for avoiding involuntary object selection is to implement dwell time (a brief delay), which helps distinguish between general viewing and voluntary object selection  \cite{hansen2018fitts, schuetz2019explanation, majaranta_2014}.
Prior studies have found that optimal dwell times vary between 150 and 1000 milliseconds (ms) depending on the user and task   \cite{majaranta2004effects, majaranta2003auditory, majaranta2012communication, chen2019using, paulus2021usability}.
Other than dwell-time, mouse-click  \cite{mouseclick}, speech  \cite{speech}, smooth pursuit  \cite{khamis2018vrpursuits}, gaze gestures  \cite{drewes2007eye}, pinch  \cite{mutasim2021pinch}, hand gestures  \cite{esteves2020comparing}, have been explored to enhance interaction accuracy.

Although our use of an offline dataset limits the incorporation of interactive techniques like mouse clicks, gaze gestures, or pinches, we will instead utilize dwell time in our study.
It allows for hands-free interaction, eliminating the need for physical actions like clicking or speaking and making it highly accessible for users with mobility impairments  \cite{impairments} or those in noisy environments  \cite{bulling2009wearable}.
Dwell-based gaze interaction has effectively reduced errors in environments with large fields of view  \cite{blattgerste2018advantages}.
Moreover, it involves straightforward thresholding based on fixation duration, faster compared to click-based interactions  \cite{sibert2000evaluation}.
This paper aims to provide a detailed analysis of spatial accuracy during simulated interaction based on dwell time.
As the optimal duration for dwell time depends on the specific requirements of the application, we will further investigate the spatial accuracy of the interaction varying the dwell times.

\section{METHODOLOGY: REAL-TIME GAZE INTERACTION SIMULATION}

\begin{figure}[htbp]
\centering
\includegraphics[width=0.95\textwidth]{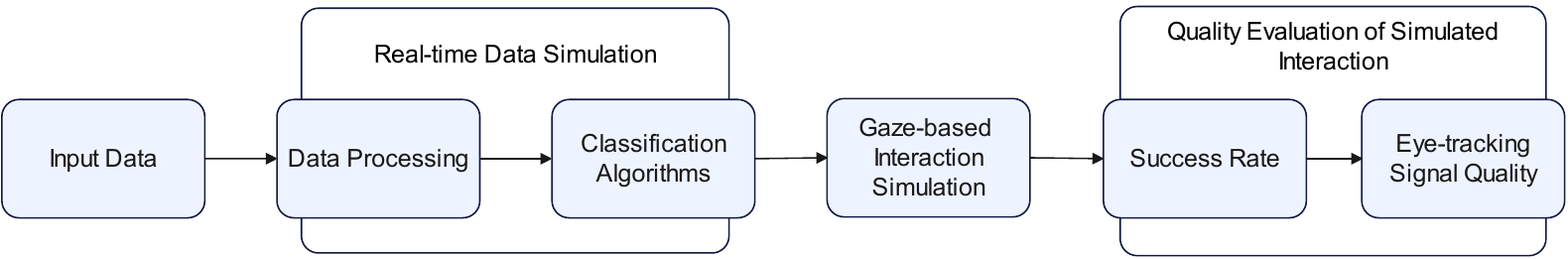}
\caption{Conceptual Diagram of the Real-time Gaze Interaction Simulation Methodology}
\label{fig:diagram}
\end{figure}

We have divided the Real-time Gaze Interaction Simulation methodology into four steps: (1) Input Data, (2) Simulating Real-time data stream, (3) Simulating Gaze-based Interaction, and (4) Evaluating the Quality of Simulated Interaction.
Fig. \ref{fig:diagram} presents a conceptual diagram of the methodology and we provide a detailed explanation of each step in the subsequent sections.

\subsection{Dataset}

The dataset used in this study is the publicly available GazeBase dataset  \cite{Gazebase}. 
Eye movement data were collected using an EyeLink 1000 (SR-Research Ltd, Ontario, Canada) eye-tracker, with a sampling rate of 1000 Hz.
The dataset comprises 12,334 monocular recordings (left eye only) from 322 college-aged participants, collected over three years across nine rounds (Round 1 to Round 9). 
Each recording includes the horizontal and vertical eye movements, measured in degrees of visual angle (dva).
Participants completed seven different eye movement tasks: random saccades (RAN), reading (TEX), fixation (FXS), horizontal saccades (HSS), two video viewing tasks (VD1 and VD2), and a video-gaming task (Balura game, BLG). Each round involved two recording sessions per subject, with a 20-minute interval between sessions. 

This study only used data from the RAN task in Round 1 (322 Subjects $\times$ 2 Sessions = 644 recordings).
The RAN task has both gaze points and target points. Target points are required for the Rank-1 fixation selection method, which will be discussed later.
During this task, participants were instructed to fixate on the center of a bull’s-eye target (diameter= one dva) that moved randomly across the display monitor. 
The target's position varied within ±15 and ±9 dva in the horizontal and vertical directions, respectively, with a minimum displacement amplitude of two dva between successive target positions. 
The target was displayed on a black background, remaining at each location for one second. 
The trajectory was randomized for each iteration, resulting in variations in the stimulus across participants, sessions, and rounds. 
The distribution of target positions was selected to ensure uniform coverage of the display area.
Readers are referred to the original publication of GazeBase dataset  \cite{Gazebase} for more details.

\subsection{Simulating Real-time Data Stream}

\subsubsection{Data Preprocessing}
In our study, we use the offline GazeBase dataset and simulate a real-time data stream, specifically from the RAN task. 
GazeBase RAN data includes timestamp, horizontal and vertical eye movement channels, and their corresponding target channels.
We started by extracting the horizontal and vertical channels from the gaze data before applying any classification algorithms.
In GazeBase, gaze position corresponds to any missing gaze samples within each recording that are specified as Not-a-Number (NaN).
The total number of missing gaze samples i.e. NaN in the GazeBase dataset can be referred to as data loss.
The GazeBase dataset has a median data loss of 1.12\% with a standard deviation (SD) of 4.47\%, while the maximum data loss observed in a recording was 48.01\%.
NaN values are appropriately handled using forward filling, considering the real-time data stream.
When a NaN value is found, it is replaced with the most recent valid data point, effectively carrying the last known value forward.
If the first value in the data stream is NaN, it is replaced with a predefined default value of 0.

\subsubsection{Eye Movement Classification Algorithms}
\label{sec:algos}
The methodology implements real-time versions of three eye movement classification algorithms—IVT, IDT, and IKF, to identify physiological fixations from each of the recordings. 

\paragraph{Velocity Threshold-based Identification (IVT)}
IVT distinguishes between fixations and saccades based on the velocity of eye movements.
We calculated the instantaneous velocity for individual channels.
The overall velocity magnitude is calculated, which combines both channels to get a value representing velocity at each time step.
Fixations are identified when the velocity of the eye movement falls below a velocity threshold, while saccades are detected when they exceed that threshold.

\paragraph{Dispersion Threshold-based Identification (IDT)}
IDT algorithm identifies fixations by calculating the dispersion of gaze points within a specified minimum temporal window (duration-threshold).
A fixation is detected if the gaze points within that minimum temporal window remain close to each other and do not exceed the dispersion threshold.
The dispersion threshold determines the maximum allowed dispersion (in dva) of gaze points for the algorithm to classify them as part of the same fixation. 
The dispersion of points in the window has been calculated following the formula from  \cite{Salvucci}, Dispersion=[max(x) - min(x)] + [max(y) - min(y)].
If the dispersion exceeds this threshold, it indicates a fixation's end.

\paragraph{Kalman Filter-based Identification (IKF)}
We implemented a Kalman filter-based approach following  \cite{Koh2009}, to estimate the state of our system from eye-tracking measurements based on the general mathematical framework described by Brown \& Hwang  \cite{brown_kalman}.
Our Kalman filter was initialized with the following parameters inspired by  \cite{komogortsev_2007kalman, Koh2009}---
The Kalman Gain is initially set to zero and updated throughout the filtering process as more data becomes available. 
The state vector is initialized to [0, 0], which represents the initial position and velocity for both channels. 
Correspondingly, the measurement vector is also initialized to $[0, 0]$, indicating that the observed measurements are initially zero.
The state covariance matrix is set as the identity matrix, signifying the initial certainty regarding the state.
To model the system, the state transition matrix assumes constant velocity and incorporates the time step ($\Delta t$) to update both position and velocity over time. 
The observation matrix is designed to directly observe the position component, linking the state and the measurements.
The measurement noise covariance is set to one, representing the expected noise level in the observations. 
Additionally, the process noise covariance matrix, which is also set as the identity matrix, accounts for uncertainties in the process, allowing the model to handle random disturbances effectively.

To detect changes in gaze behavior, we calculated the cumulative sum of the squared differences between the predicted and measured values from the Kalman filter, normalized by the deviation. 
These values were used as thresholds to detect fixations.
Specifically, the difference between chi-squared values over a sliding window was computed to identify changes in gaze behavior. 
If the chi-squared value difference was below a defined threshold for each data point, the point was classified as part of a fixation.

\subsubsection{Parameters Selection based on Friedman-Komogortsev Method}
\label{subsec:threshold}

Fixation labeling by the Friedman-Komogortsev Method (FKM) classification algorithm\footnote{FKM is an improved version of the MNH algorithm  \cite{FKM}, available at \url{https://hdl.handle.net/10877/16339}} has been used as ground truth for selecting the optimal parameters for the classification algorithms employed above. 
The best parameter set has been selected based on the highest classification accuracy, with FKM labeling as the ground truth.
We implemented algorithms with the best parameters--- 

(1) IVT: Implemented with a velocity threshold of 30 deg/sec (tested with thresholds 20, 30, and 40 deg/sec).

(2) IDT: Implemented with a dispersion threshold of 0.5 deg and a minimum duration of 30~ms 
(tested with dispersion thresholds of 0.5, 0.75, 1.0 deg/sec and minimum durations ranging from 20 to 60~ms in 10~ms intervals).

(3) IKF: Implemented with a chi-square threshold of 3.75, a window size of 5, and a deviation of 1000 (tested with chi-square thresholds from 1 to 5.75 in 0.25 increments, window sizes of 3,5,7 and deviations from 500 to 2000 in 500 increments).

\subsection{Simulating Gaze-based Interaction}

As the GazeBase dataset is not interaction-based, we make the assumption that allows to use of it for simulation purposes. 
We are using the RAN task which has both gaze positional values and targets. 
In our simulation, these targets are treated as the objects with which participants intend to interact.

Rank-1 Fixation Selection is our proposed method to define the period when interaction is made.
Starting with the RAN data from GazeBase Round-1, we have identified all fixation points relative to each target using the classification algorithms mentioned in Section \ref{sec:algos}.
Dwell time will be our interactive technique.
Based on the dwell time threshold, we will define whether is there an interaction there or not.
Although previous research has suggested that a dwell time range of 150 to 1000~ms is appropriate  \cite{majaranta2004effects, majaranta2003auditory, majaranta2012communication, chen2019using, paulus2021usability, chen2021adaptive, ware1986evaluation, schuetz2019explanation, miniotas2006speech}, the GazeBase dataset presents a challenge: while the target remains at a specific coordinate for 1000~ms only, it is practically impossible to find fixation periods of 1000~ms in this dataset, using the specific event classification method that we employ.
So, higher dwell time is not feasible in the study.

As the human fixation duration typically ranges from 150 to 300~ms  \cite{fixation_duration}, in this study, we compared the performance of algorithms using chosen dwell times ranging from 100~ms to 300~ms.
We understand 100~ms is too low for practical use, however, employ this theoretical lower limit to study the eye tracking signal quality  \cite{vertegaal2008fitts}. 
For each target, we calculated the Euclidean distance in dva between the target and all identified fixation periods.
As we are using a sliding window technique to find the interaction, there is a possibility of finding multiple fixation periods of a certain duration (meeting the dwell time threshold).
First, we identify all fixation periods and then we choose the one closest (the least distance) to target.
We termed it as the Rank-1 Fixation.
So, Rank-1 Fixation represents the fixation period that exhibited the lowest Euclidean distance to the target, indicating the closest spatial proximity which means the fixation period exhibits the best spatial accuracy for the given target.
Saccades generally exhibit a latency of about 200~ms  \cite{leigh2015neurology} before being initiated in response to a new stimulus or target. 
During this delay, the eyes may continue to fixate on the previous target while the brain processes and prepares for the upcoming saccadic movement. 
As a result of this saccadic latency, a fixation that began on a previous target may still be ongoing when the new target appears. 
This introduces the possibility that some fixations recorded on the new target might be carryovers from the prior target; hence, the spatially closest fixation was considered.

\begin{figure}[htbp]
\centering
\includegraphics[width=0.7\textwidth]{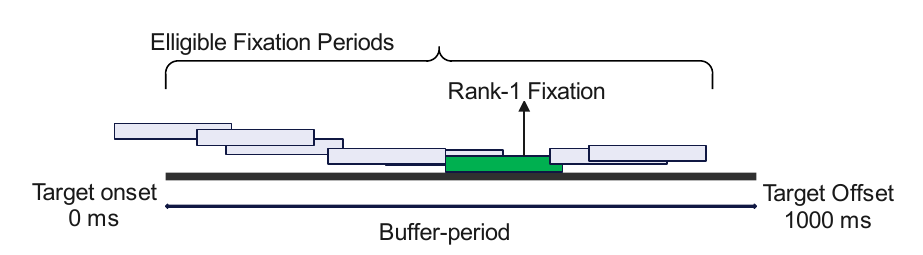}
\caption{Rank-1 Fixation Selection. Green represents the Rank-1 Fixation (Trigger-event).}
\label{fig:rank1}
\end{figure}

In the context of our study, a ``trigger-event'' refers to the moment where the user is most likely looking directly at the target, making a gaze-based interaction triggering an event that will be analyzed.
On the other hand, ``buffer-period'' is the time allowed to make the trigger-event. 

The Rank-1 Fixation occurring within the preset buffer-period after the target's onset has been considered the ``trigger-event'' for that target.
Fig. \ref{fig:rank1} provides a visual demonstration of how Rank-1 Fixation has been identified from all eligible fixation periods.
In this study, primarily the trigger-event was made within one second following the target's onset.
That means primarily the buffer-period is one second.
In the latter of the paper, we also compare performance with variable buffer-period (400~ms to 1000~ms). 
Evaluation of varying buffer-period for trigger-events is significant because it directly affects the spatial accuracy and reliability of eye-tracking systems in real-time applications.
The longer the buffer-period, the more data is available to define the trigger-event.

\subsection{Evaluating Quality of Simulated Interaction}

In our study, we will evaluate 
(1) the effectiveness of our methodology in defining trigger events under different conditions, including variations in classification algorithms, dwell times, and buffer-periods. 
(2) the signal quality of these trigger events under the same conditions.

\subsubsection{Effectiveness in Defining Trigger-event}

In this study, we evaluate the effectiveness of our simulation methodology for defining the trigger-event by calculating the success rate.
Success rate is based on the percentage of trigger-events that were successfully identified out of the maximum number of trigger-events possible.
In our study, we analyze success rates using the GazeBase RAN task, involving 322 participants who each completed two sessions. 
This resulted in a total of 644 recordings. 
Each recording includes 100 targets, cumulating in 64,400 targets across all recordings. 
For our interaction simulation methodology, we assume one trigger-event per target, setting the potential maximum at 64,400 trigger-events.

The success rate is defined as follows:
\begin{equation*}
    \text{Success rate}_i = \frac{\text{Number of defined trigger-events for recording i}}{\text{Total number of targets for each recording = 100}} \tag{1}
\end{equation*}

\begin{equation*}
    \text{Success rate} = \frac{\sum_{i=1}^{R} \text{Success rate}_{i}}{\text{Total number of recordings (R=644)}} \tag{2}
\end{equation*}

It is important to note that a simulation methodology with specific conditions (such as \textit{X} classification algorithm, \textit{Y} dwell time and \textit{Z} buffer-time), may not define trigger-event for each target. 
We will use the success rate to evaluate the effectiveness of our simulation methodology in defining trigger-events under different conditions. 
This metric reflects the overall success of our methodology across all recordings; a higher success rate indicates better performance.

\subsubsection{Eye-tracking Signal Quality}

We have calculated the spatial accuracy, indicative of the eye-tracking signal's quality, for trigger events linked to all targets within the eye movement recordings. 
Spatial accuracy is referred to by various terms and calculated through different methods in prior eye-tracking research  \cite{holmqvist2012eye, hornof2002cleaning, lohr2019evaluating}.
In our research, spatial accuracy is defined as the angular offset between the center of the target and the centroid of the participant’s gaze vector during the corresponding fixation. 
For each trigger event, spatial accuracy is computed according to the following equation:

\begin{equation}
\theta =  \frac{180}{\pi} \arccos\left( \frac{G \cdot T}{\|G\| \|T\|} \right) \tag{3}
\end{equation}

where, 
$G$ is the centroid of the gaze-position vector, $T$ is the center of the target vector, and $\theta$ is the angular distance in degrees of visual angle.

We present our results on spatial accuracy using the eye-tracking signal quality descriptors from \cite{U|E_Sam}: U50|E50 and U95|E95. 
These descriptors represent a combination of user-centric and population-centric evaluations, providing insights into spatial accuracy measured at trigger-event at both the individual and population levels. 
``E'' represents individual-level error percentiles, indicating the distribution of spatial accuracy for each individual. 
``U'' represents user percentiles, which describe population-level spatial accuracy by aggregating spatial accuracy across different individual error percentiles.

Specifically, E50 and E95 indicate the distribution of spatial accuracy for an individual user, where E50 represents the median error and E95 represents the worst-case error with the 95th percentile of error.
On the other hand, U50, and U95 represent population-level insights aggregated across the respective E percentiles of individual users. 
Thus, U50|E50 reflects the median value of E50 spatial accuracy values observed across the population during the trigger-event.
Finally, U95|E95 describes the 95$^{th}$ percentile of E95 error observed across the population in their worst-case interaction scenarios.

\section{RESULTS}

\subsection{Success Rate of the Simulation Methodology under Varying Constraints}
\label{subsec:sucess_rate}

Success rate is based on the percentage of trigger-events that were successfully identified out of the 64,400 trigger-events that we assume to be present.
Table \ref{tab:fix_acc} presents the success rate in defining the trigger-event for three eye movement classification algorithms--- IVT, IDT, and IKF under varying dwell time and buffer-period conditions\footnote{Extended version of the Table \ref{tab:fix_acc} is added in the supplementary materials.}. 
Performance is assessed based on the following conditions:

\begin{enumerate}
    \item 
    \textit{Varying dwell time with a fixed buffer-period of 1000~ms}--- 
    As shown in Table \ref{tab:fix_acc1}, the dwell time was varied while the buffer-period remained fixed at 1000~ms. 
    The IDT algorithm achieved a higher success rate in defining trigger-event compared to others at dwell times of 200~ms and less. 
    Conversely, IDT performed the worst among the implemented algorithms for a dwell time of 300~ms.
    The IKF algorithm performed better compared to others for longer dwell times (250~ms and 300~ms).
    The IVT algorithm showed lower performance, particularly at longer dwell times. 
    
    The results highlight that shorter dwell times generally yield a higher success rate in defining trigger-event, with IKF showing the highest resilience across different dwell time settings.

    \item 
    \textit{Varying buffer-period with a fixed dwell time of 100~ms}---
    As shown in Table \ref{tab:fix_acc2}, the buffer-period period was varied while maintaining a fixed dwell time of 100~ms. 
    The IDT algorithm maintained the highest overall accuracy, exceeding 99\% accuracy for longer buffer-periods: 800~ms and 1000~ms. 
    IKF performed comparably to IDT but experienced a slight decline as buffer-period decreased, achieving 91.7\% at 400~ms.
    The IVT algorithm showed the worst accuracy across all buffer-periods.
    
    Longer buffer periods provide more data, benefiting IDT, which relies on spatial thresholds, more than others.
    On the other hand, IVT is more sensitive since it depends on velocity, which fluctuates with noise.
    IKF’s predictive filtering helps handle noise and missing data, but it struggles with shorter periods due to insufficient data for accurate state updates.
    Results indicate that IDT is less affected by noise as it uses spatial threshold.

\end{enumerate}

\begin{table}[ht!]
\centering
\caption{Comparison of success rate(\%) between algorithms for various dwell time and buffer-period. Higher values indicate better performance.}
\label{tab:dt_lat}
\begin{subtable}[t]{0.48\textwidth}
    \centering
    \caption{Dwell times varies with fixed buffer-period = 1000~ms.}
    \label{tab:dt}
    \begin{tabular}{cccc}
    \hline
    \multirow{2}{*}{Dwell time (ms)} & \multicolumn{3}{c}{Algorithms}\\ 
    \cline{2-4} 
             & IVT   & IDT   & IKF   \\ \hline
    300      & 64.2  & 56.3  & 75.3 \\ \hline
    250      & 71.1  & 73.2  & 82.5 \\ \hline
    200      & 78.2  & 90.2  & 89.6 \\ \hline
    150      & 83.4  & 97.2  & 94.1 \\ \hline
    100      & 87.8  & 99.4  & 97.2  \\ \hline
    \end{tabular}
    \label{tab:fix_acc1}
\end{subtable}%
\hfill
\begin{subtable}[t]{0.48\textwidth}
    \centering
    \caption{Buffer-period periods varies with fixed dwell time = 100~ms.}
    \label{tab:buffer-period}
    \begin{tabular}{cccc}
    \hline
    \multirow{2}{*}{Buffer-period (ms)} & \multicolumn{3}{c}{Algorithms} \\ 
    \cline{2-4} 
          & IVT   & IDT   & IKF   \\ \hline
    1000  & 87.8  & 99.4  & 97.2  \\ \hline
    800   & 86.9  & 99.1  & 96.8 \\ \hline
    600   & 84.5  & 97.9  & 95.6 \\ \hline
    400   & 80.7  & 94.8  & 91.7 \\ \hline
    \end{tabular}
    \label{tab:fix_acc2}
\end{subtable}
\label{tab:fix_acc}
\end{table}

Now let us explain with an example. For simulation with IVT, dwell time of 100~ms and buffer-period of 1000~ms, we have a success rate of 87.8\% from Table \ref{tab:fix_acc1}.
A success rate of 87.8\% corresponds to 64,400 $\times$ 0.878 $\equiv$ 56,543  defined trigger-events and 64,400-56,543 = 7,857 targets with no trigger-event.
It indicates that the methodology with IVT, dwell time of 100~ms, and buffer period of 1000~ms fails to define trigger events for the 7,857 targets across recordings.
It is important to consider that this failure reflects the algorithm's inability to detect physiological fixations corresponding to each target. 
This performance limitation holds regardless of the interaction technique employed for gaze interaction.

\subsection{Comparison of Spatial Accuracy of Trigger-event under Varying Constraints}

\subsubsection{Varying Classification Algorithms}
\label{subsec:result1}

\begin{figure*}[htbp]
\centering
\includegraphics[width=\textwidth]{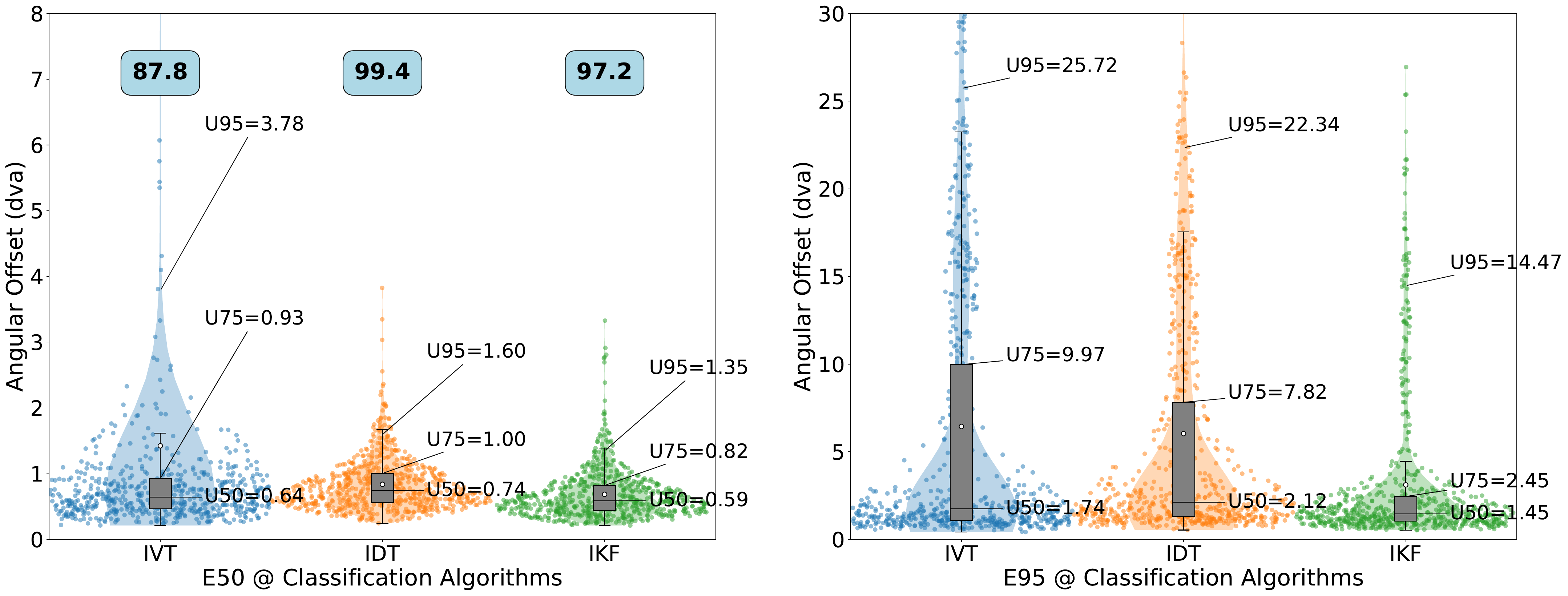}
\caption{Distribution of E50 (left-plot) and E95 (right-plot) spatial accuracy of trigger-event across the user population for three different classifications. White circles represent the mean of their respective distributions. Population-centric percentile for the respective user error percentile is annotated in the figure. Success rates (\%) are annotated on the left plots. The ordinate on the left plot is scaled for better visualization. Dwell time=100~ms and Buffer-period=1000~ms.}
\label{fig:comparison_algo_100}
\end{figure*}

In this analysis, we calculate the spatial accuracy of the trigger-event for each recording thrice using three different classification algorithms within the simulation methodology.
Then we compare the calculated spatial accuracy between algorithms using violin plots, as shown in Fig. \ref{fig:comparison_algo_100}.

We conduct this analysis using a dwell time of 100~ms and buffer-period of 1000~ms because these conditions produced the highest success rates for defining trigger-events for each algorithm (Section \ref{subsec:sucess_rate}).
Success rates for accurately defining the trigger-events (\%) are annotated on the left plots for each simulation setting.
The respective classification algorithms use the threshold parameters from Section \ref{subsec:threshold}.
The left plot shows the distribution of E50 spatial accuracy for trigger-events across the user population for three different classifications. 
The right plot displays the distribution of E95 spatial accuracy for trigger-events across the user population for the same classifications.
Across all three algorithms, the angular offsets are relatively small for the distribution of E50 spatial accuracy, with IKF having the smallest median spatial accuracy (U50|E50 = 0.59 dva) compared to IVT and IDT.
IVT has a higher 95th percentile value (U95|E50 = 3.78 dva) than IDT and IKF, suggesting that it may have larger outliers, even though the median performance(U50|E50) is close across methods.
For the distribution of E95 spatial accuracy, all angular offsets are significantly larger than those of E50. 
IVT has the largest spread, with U95|E95 = 21.26, indicating many high offset values.
IDT also shows a large spread but with slightly lower values than IVT.
IKF continues to perform better with lower median and upper quartile values (U50|E95 = 1.45 dva, U75|E95 = 2.45 dva), but its U95|E95 (14.47 dva) is still quite large.

\subsubsection{Varying the Dwell Time}
\label{subsec:result2}

\begin{figure*}[htbp]
\centering
\begin{subfigure}[b]{\textwidth}
    \centering
    \includegraphics[width=\textwidth]{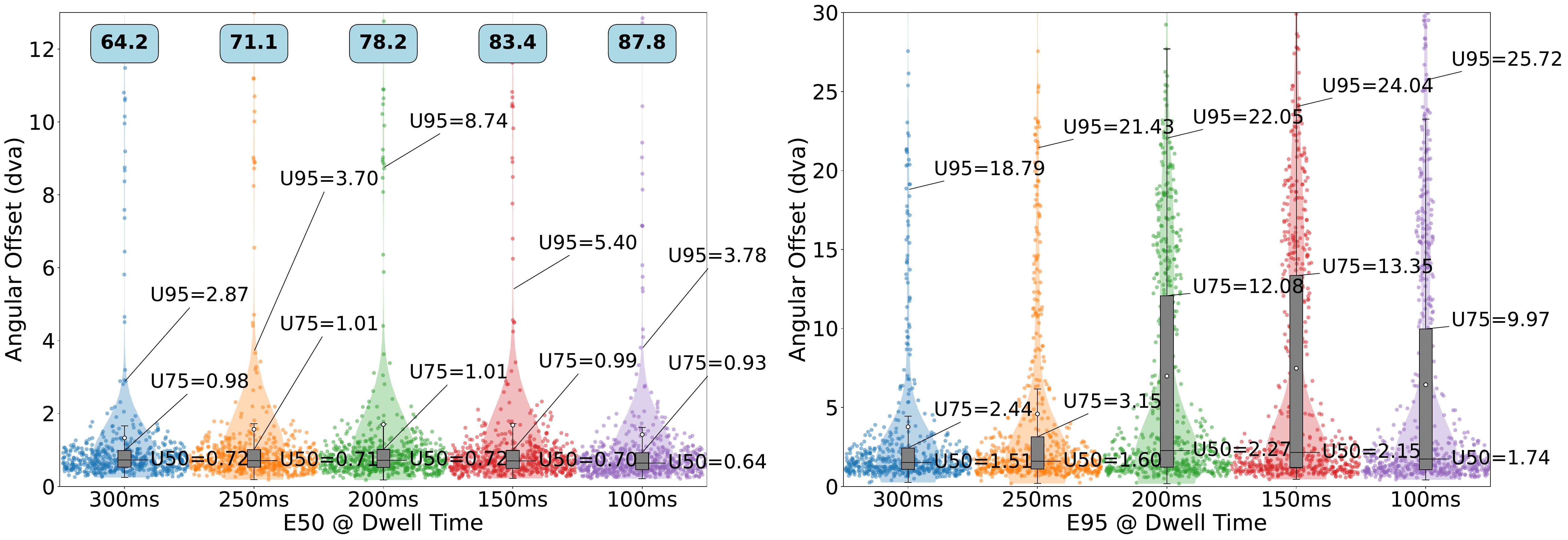}
    \caption{IVT}
    \label{fig:comparison_dt_ivt}
\end{subfigure}
\vspace{0.1cm}

\begin{subfigure}[b]{\textwidth}
    \centering
    \includegraphics[width=\textwidth]{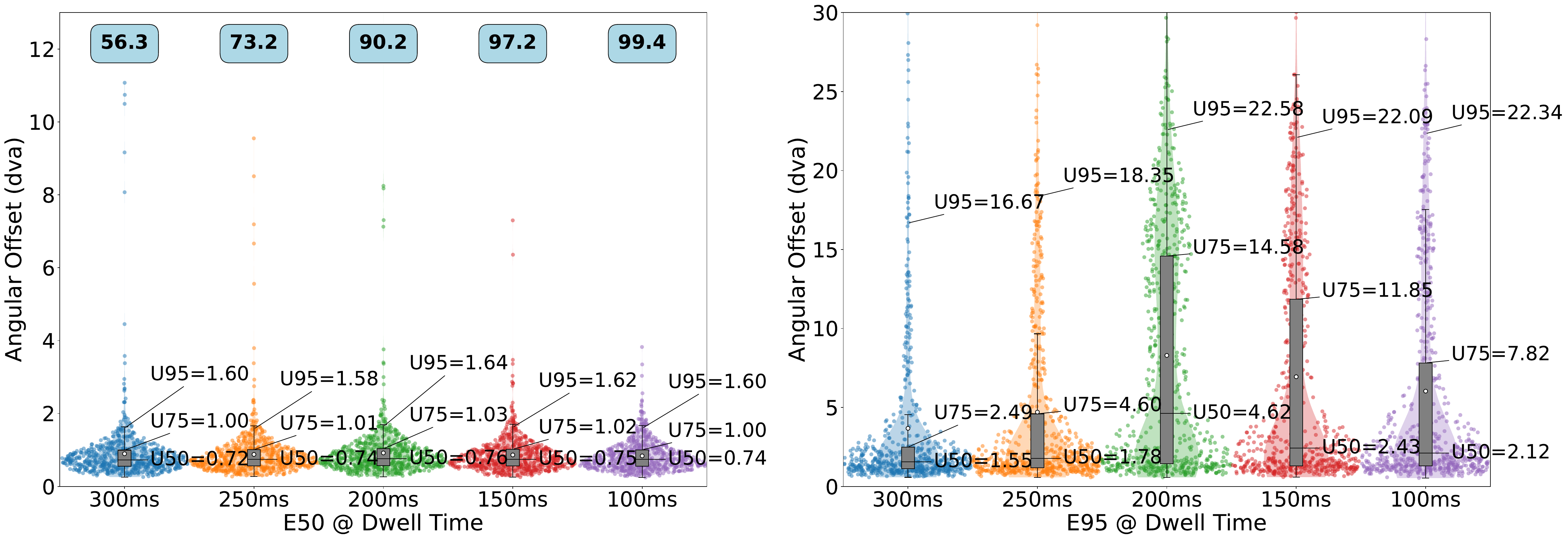}
    \caption{IDT}
    \label{fig:comparison_dt_idt}
\end{subfigure}
\vspace{0.1cm}

\begin{subfigure}[b]{\textwidth}
    \centering
    \includegraphics[width=\textwidth]{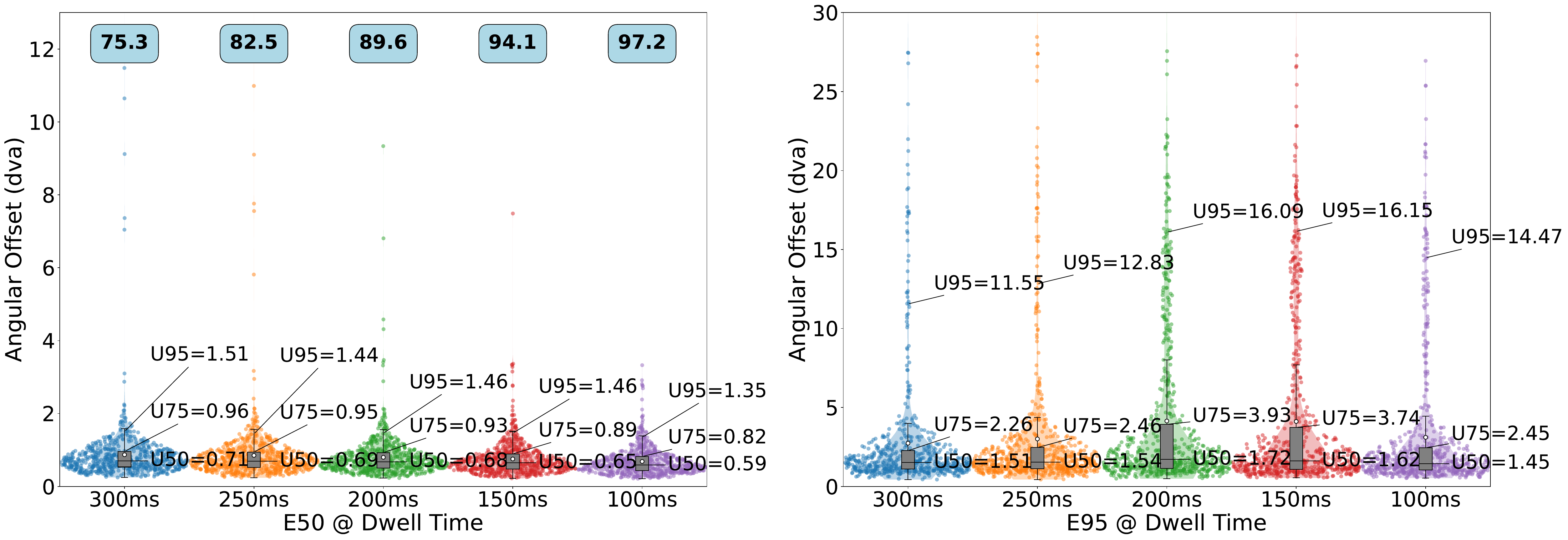}
    \caption{IKF}
    \label{fig:comparison_dt_ikf}
\end{subfigure}

\caption{Comparison of the distribution of E50 and E95 spatial accuracy of trigger-events for IVT, IDT, and IKF across the user population, with varying dwell times. U50, U75, and U95 respective to their individual-level error percentiles are annotated in the plots. Success rates (\%) are annotated on the left plots. The ordinate on the left plots is scaled for better visualization. Buffer-period = 1000~ms.}
\label{fig:comparison_dt_all}
\end{figure*}

In Fig. \ref{fig:comparison_dt_all}, we compare the spatial accuracy of trigger-events for the (a) IVT, (b) IDT, and (c) IKF algorithms, across different dwell times ranging from 100~ms to 300~ms. 
A fixed buffer-period of 1000~ms is used in this comparison.
Success rates for accurately defining the trigger-events (\%) are annotated on the left plots for each simulation setting.
The left plot displays the E50 spatial accuracy distribution, while the right plot shows the E95 spatial accuracy distribution, both presented as violin plots for each algorithm.
As dwell time decreases, there is a general downward trend in the U50, U75, and U95 values. 
Specifically, with the 100~ms dwell time, the lowest angular offset is observed: 0.64 dva for IVT, 0.74 dva for IDT, and 0.59 dva for IKF, indicating that shorter dwell times lead to superior spatial accuracy. 
Notably, the IKF algorithms consistently outperform the other algorithms across all metrics, as annotated in the plot.
This phenomenon is common for both the E50 and E95 distributions.

\subsubsection{Varying the Buffer-period}
\label{subsec:result3}

\begin{figure*}[htbp]
\centering
\begin{subfigure}[b]{\textwidth}
    \centering
    \includegraphics[width=\textwidth]{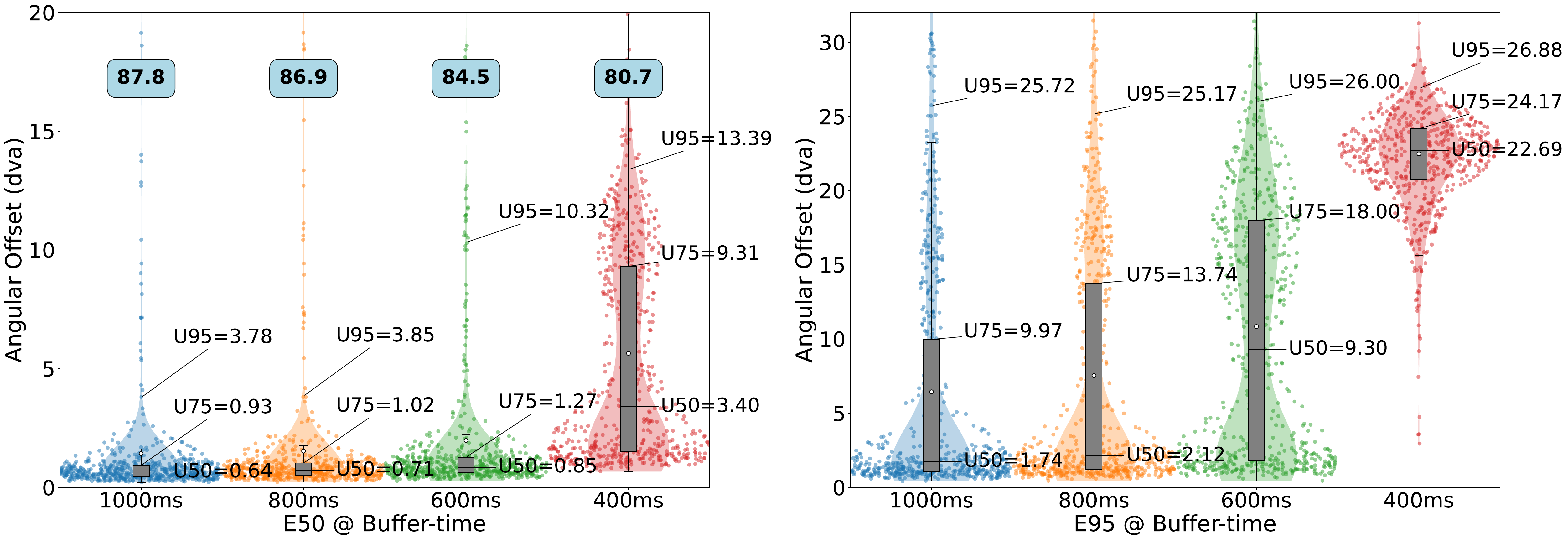}
    \caption{IVT}
    \label{fig:comparison_ce_ivt}
\end{subfigure}
\vspace{0.1cm}

\begin{subfigure}[b]{\textwidth}
    \centering
    \includegraphics[width=\textwidth]{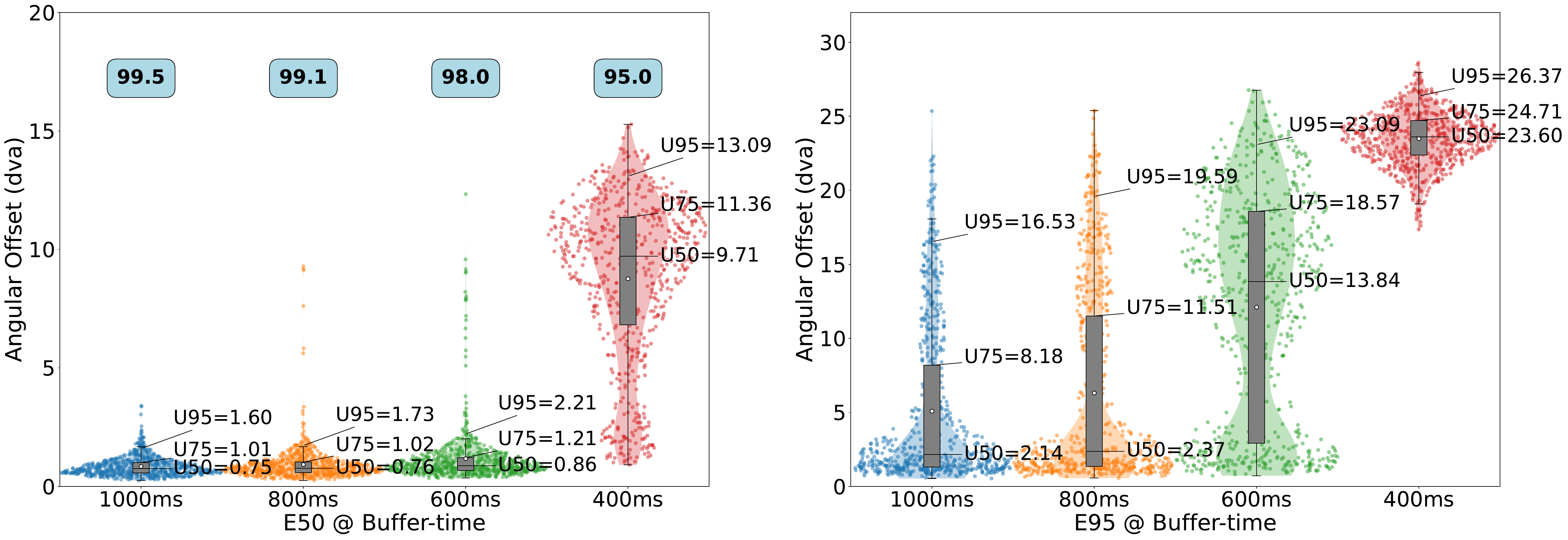}
    \caption{IDT}
    \label{fig:comparison_ce_idt}
\end{subfigure}
\vspace{0.1cm}

\begin{subfigure}[b]{\textwidth}
    \centering
    \includegraphics[width=\textwidth]{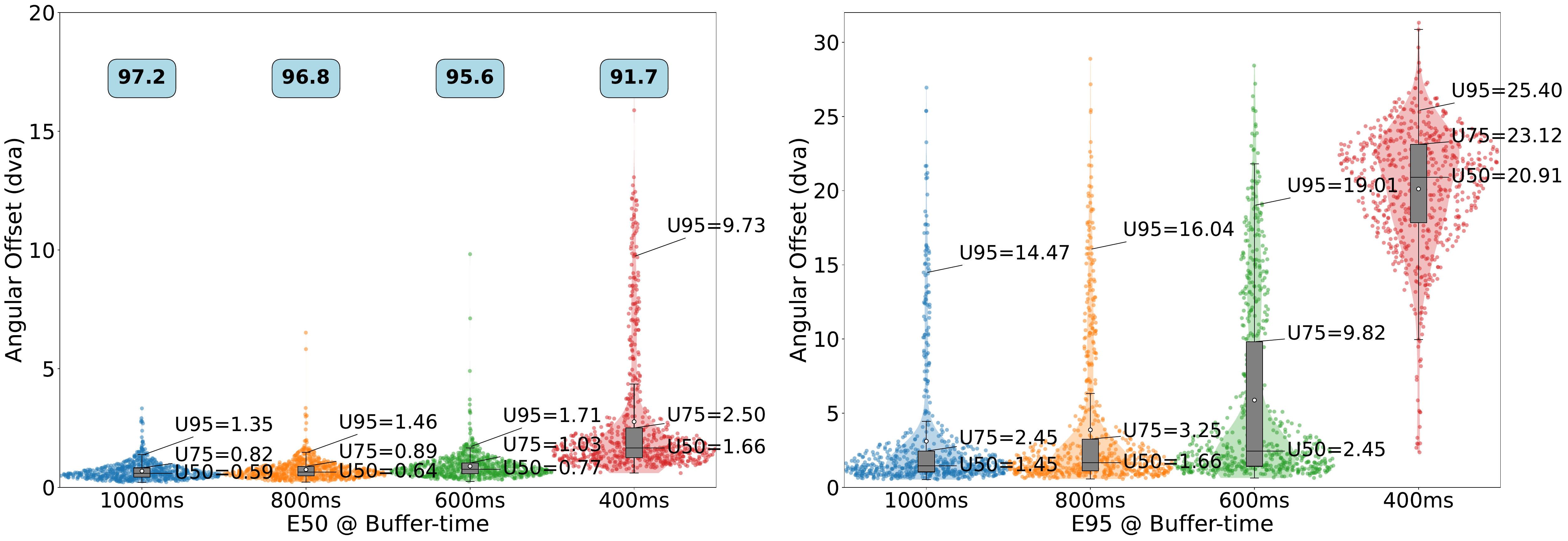}
    \caption{IKF}
    \label{fig:comparison_ce_ikf}
\end{subfigure}

\caption{Comparison of the distribution of E50 and E95 spatial accuracy of trigger-events for IVT, IDT, and IKF across the user population, with varying buffer-period. U50, U75, and U95 respective to their individual-level error percentiles are annotated in the plots. Success rates (\%) are annotated on the left plots. The ordinate on the left plot is scaled for better visualization. Dwell time = 100~ms.}
\label{fig:comparison_ce_all}
\end{figure*}

In the above-presented results in Section \ref{subsec:result1} and \ref{subsec:result2}, we have used the fixed buffer-period of 1000~ms. 
In Fig. \ref{fig:comparison_ce_all}, we compare the spatial accuracy of trigger-events for the (a) IVT, (b) IDT, and (c) IKF algorithms, across different buffer-periods from 400~ms to 1000~ms. 
A dwell time of 100~ms is used in this comparison.
Success rates for accurately defining the trigger-events (\%) are annotated on the left plots for each simulation setting.
For each algorithm, the left plot displays the E50 spatial accuracy distribution, while the right plot shows the E95 spatial accuracy distribution, both presented as violin plots.
As buffer-period shortens, there is a general upward trend in the U50, U75, and U95 values.
Besides, when the buffer-period goes down to 400~ms, all  IKF performs significantly better than the other two.
Specifically, with a buffer-period of 1000~ms, the lowest angular offset is observed: 0.64 dva for IVT, 0.75 dva for IDT, and 0.59 dva for IKF.
Plots indicate the longer the buffer-period, the better the spatial accuracy.
These plots demonstrate that reducing buffer-period negatively impacts performance, particularly in terms of spatial accuracy.
The IKF algorithms consistently outperform the others across both the E50 and E95 distributions.

Though we have analyzed varying the buffer period, the actual trigger-event onset is much quicker compared to the buffer-period available in a particular simulation setting. 
For instance--- with IKF, dwell time of 100~ms, buffer-period of 1000~ms, the median actual trigger-event onset is 592~ms  with an SD of 224~ms---quicker than that of IVT (604~ms) and IDT (810~ms).
Detailed results are added in supplementary materials.

\section{DISCUSSION}


\subsection{Implications of Classification Algorithms}

The findings from Table \ref{tab:fix_acc} show that IDT outperforms other algorithms in accurately defining trigger-events across all buffer-periods and for shorter dwell time.
IKF is also effective in accurately defining trigger-events, even better than IDT for longer dwell time scenarios such as 250~ms and 300~ms.
Conversely, our spatial accuracy-based analyses show that the IKF algorithm consistently selects Rank-1 Fixations that produce better spatial accuracy, minimizing angular offset, which enhances the spatial accuracy of trigger-event. 
Although its advantage in the E50 distribution is modest compared to IDT in some cases, IKF shows substantial improvement in the E95 distribution, particularly in handling outliers in challenging conditions, showing notably superior performance in U75|E95. 
Spatial accuracy using the IKF also improves as dwell time shortens. 
Moreover, IKF outperforms others consistently in spatial accuracy while varying buffer-period.
Our findings have important implications for both algorithm selection and the development of real-time gaze-based interfaces.
As findings suggest that no single algorithm suits all, the trade-off between spatial accuracy and success rate emphasizes the importance of choosing the right algorithm based on specific demands.
The results demonstrate that the IKF algorithm might be a good choice for applications requiring precise gaze interaction such as gaze-based interfaces, and virtual environments.
Conversely, the IDT algorithm offers a more reliable solution for applications that prioritize accurately defining trigger-events, especially under temporally constrained conditions. 
The significant limitations of IVT, indicate that it may not be suitable for applications that require both precise gaze interaction and temporal adaptability.

\subsection{Implications of Dwell Time}

Irrespective of the classification algorithms utilized in the study, our analysis indicates that success rate in defining trigger-event tends to improve with shorter dwell times, particularly noting a significant improvement when reducing from 250~ms to 200~ms.
However, the results illustrated in Figure. \ref{fig:comparison_dt_all} could be misleading.
This is because spatial accuracy should improve with shorter dwell times as we defined trigger-event from the best set of data points compared to the target, although this is not consistently the case.
It is important to recognize that spatial accuracy in our study is calculated solely based on trigger-events and that these trigger events might be defined differently depending on which classification algorithm was used to classify the data.
Consequently, longer dwell times can result in poor performance in accurately defining trigger-events, leading to an unreliable comparison between different dwell times due to the varying number of trigger-events.
Therefore, success rate should be the first to look at when varying the dwell time, but not the only one to rely on.
Though shorter dwell time yields better performance in accurately defining trigger-event, it will increase the likelihood of involuntary interaction, Midas touch problem  \cite{Midas_touch1, Midas_touch2}.

\subsection{Implications of Buffer-period}

When employing a longer buffer-period, improvements in both success rate and spatial accuracy are observed across different classification algorithms.
The underlying reason is straightforward: extending the buffer-period allows for more data to be collected, increasing the chances that a suitable trigger-event is present.
However, this also introduces a longer time delay from the target onset to the trigger-event, which can be a critical consideration for real-time applications.
This trade-off between improved spatial accuracy and time delay should be application-specific in practical settings.
Applications requiring precise eye-movement control or deliberate gaze selections benefit from a longer buffer-period for enhanced spatial accuracy. 
On the other hand, applications that involve quick interactions or use gaze for navigation may perform better with a shorter buffer-period to minimize delays.

\subsection{Source of High Angular Offset}

In our analysis, we have seen some higher angular offsets such as 26.37 dva, 26.88 dva, etc in Fig. \ref{fig:comparison_algo_100}-\ref{fig:comparison_ce_all}.
This phenomenon of poor spatial accuracy has been observed in previous studies as well  \cite{lohr2024establishing, U|E_Sam}, though it was investigated using different eye-tracking data.
We investigated the sources of such higher errors, particularly in the U95|E95 scenarios, identifying two primary causes.
First, occasional failures in eye-tracker performance led to NaN values in gaze data. 
Our method of handling these NaNs by referencing earlier values inadvertently introduced larger errors. 
Second, a consistent deviation was observed between the gaze and target positions from the beginning of the recording, likely due to a calibration error. 
For illustrative purposes, we have included additional figures as supplementary materials.

\subsection{Limitation and Future Research}

We used the RAN task from the GazeBase dataset in our study. 
The RAN task did not explicitly involve gaze-based interaction. 
Participants were instructed to focus on the jumping target. 
So, user behavior may differ when they are allowed to interact freely.
Another limitation of this study arises from the constraints of the GazeBase dataset, which limits target fixation to 1000~ms.
As a result, we could only evaluate algorithm performance with buffer-periods up to 1000~ms.
This restriction limits our ability to examine how these algorithms perform at buffer-periods exceeding 1000~ms, which could be relevant in some real-time applications with longer response times.
However, with this buffer period, we achieve a comparable spatial accuracy for trigger-events (U50|E50 with IKF = 0.59 dva) to the spatial accuracy reported in previous studies using the EyeLink 1000 eye tracker  \cite{ehinger2019new, lohr2019evaluating}.
Lastly, it is to be noted that, this paper alters neither the size of the target nor the distance between targets.
Variability in these parameters is beyond the scope of the study.

Future research will investigate filtering techniques to reduce noise and irrelevant information, assessing their impact on the spatial accuracy of the trigger-event. 
These results will serve as a baseline for evaluating how filtering affects the signal quality of the defined trigger-events.

\section{CONCLUSION}

This study developed a methodology for simulating real-time gaze interaction using the offline GazeBase dataset as a real-time data stream, evaluating three eye movement classification algorithms: IVT, IDT, and IKF.
We have introduced Rank-1 fixation selection method to define trigger-events in the gaze-interaction simulation.  
Our findings show that the IKF algorithm outperformed others in spatial accuracy, particularly for the 95$^{th}$ percentile user-level (E95) distribution, and benefited from shorter dwell times and longer buffer-periods. 
While the IDT algorithm slightly surpasses the IKF in accurately defining trigger-events during shorter dwell times, the IKF performs better with longer dwell times.
Moreover, our analysis reveals that longer buffer-periods generally yield better performance in both success rate in defining trigger-events and spatial accuracy of trigger-events. 
However, these extended buffer-periods also result in delays from the target onset to the trigger-event, which can crucial in real-time applications. 
Variability in performance across individuals underscores the importance of choosing the right algorithm, as median user error rates were similar across the three algorithms, but spatial accuracy varied significantly among the broader population.

\bibliographystyle{unsrt}
\bibliography{reference.bib}

\end{document}


\title{Supplementary Materials for \textit{Evaluating Eye Tracking Signal Quality with Real-time Gaze Interaction Simulation}}

\author{ 
{\hspace{1mm}Mehedi Hasan Raju}\thanks{corresponding author} \\
Texas State University\\
San Marcos, Texas, 78640, USA\\
\texttt{m.raju@txstate.edu} \\
\And
{\hspace{1mm}Samantha~Aziz} \\
Texas State University\\
San Marcos, Texas, 78640, USA\\
\texttt{sda69@txstate.edu} \\
\And
{\hspace{1mm}Michael J. Proulx} \\
University of Bath\\
Bath, Somerset BA2 7AY, UK \\
\texttt{m.j.proulx@bath.ac.uk} \\
\And
{\hspace{1mm}Oleg V. Komogortsev} \\
Texas State University\\
San Marcos, Texas, 78640, USA\\
\texttt{ok11@txstate.edu} \\
}
\date{}

\maketitle
\maketitle

\renewcommand{\thetable}{\arabic{table}}
\renewcommand{\tablename}{Supplementary Table}
\captionsetup[figure]{labelformat=empty}

\section{Actual Trigger-Event Onset}
\label{sup_sec:actual_trigger_onset}

We have analyzed varying the buffer period, the actual trigger-event onset is much quicker compared to the buffer-period available in a particular simulation setting.
Among the classification algorithms employed, the IKF demonstrates the fastest response across all dwell time conditions shown in the following table.
The IVT algorithm follows, exhibiting a slightly slower response than IKF. 
In contrast, the IDT algorithm shows the slowest response when comparing the median onset times of the actual trigger-event.
It is important to note that the response time varies based on the available buffer-period, experimental set-up, and applications \cite{miniotas2006speech, schuetz2019explanation, selection_time,ware1986evaluation}.
Detailed results are shown in Supplementary Table \ref{tab:dt} with variable dwell time and fixed buffer-period of 1000 milliseconds (ms).

\begin{table}[ht!]
    \centering
    \caption{Median actual trigger-event onset (standard deviation). All the numbers represent time in ms.}
    \label{tab:dt}
    \begin{tabular}{cccc}
    \hline
    \multirow{2}{*}{Dwell time (ms)} & \multicolumn{3}{c}{Algorithms}\\ 
    \cline{2-4} 
             & IVT    & IDT    & IKF   \\ \hline
    300      & 577 (147.04)  & 594 (130.54)  & 572 (147.24)  \\ \hline
    250      & 595 (168.50)  & 629 (152.73)  & 587 (167.07)  \\ \hline
    200      & 599 (202.46)  & 671 (197.01)  & 593 (194.64) \\ \hline
    150      & 603 (217.57)  & 737 (177.55)  & 594 (211.26) \\ \hline
    100      & 604 (226.94)  & 810 (149.31)  & 592 (224.35) \\ \hline
    \end{tabular}
\end{table}

\section{Source of High Angular Offset}
\label{sup_sec:high_angular_offset}

Though we investigated the sources of such higher errors, particularly in the U95E95 scenarios, identifying two primary causes.

\begin{figure*}[htbp]
\centering
\includegraphics[width=\textwidth]{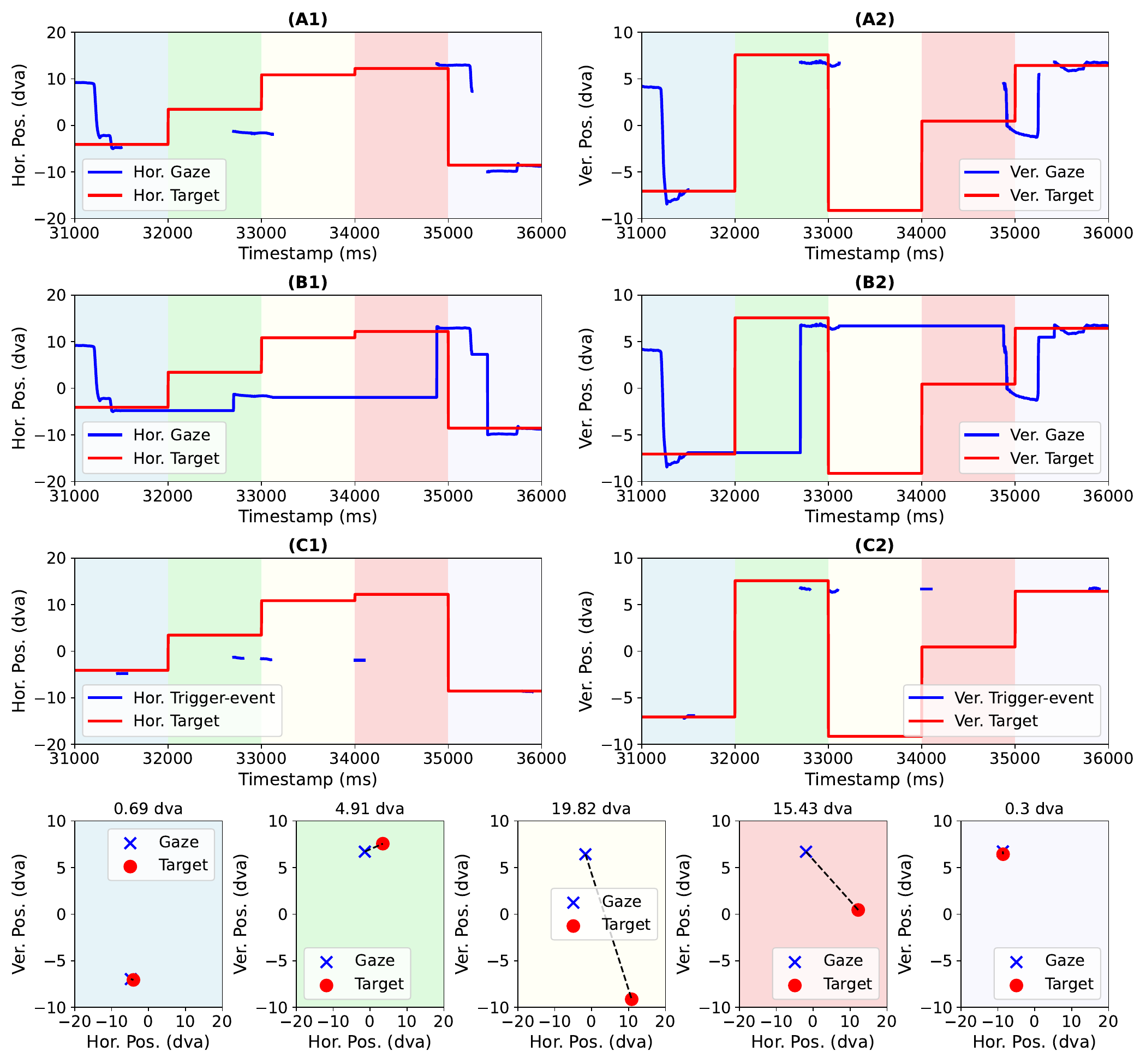}
\caption{Supplementary Fig. 1. Exemplar: Angular offset in trigger-events for noisy eye-tracking data. 
(A1-A2) Display horizontal and vertical channels of an eye-tracking signal from GazeBase RAN data containing NaNs, with five different targets represented in separate colors.
(B1-B2) Illustrate the same signal as in (A1) and (A2) but with NaNs replaced (if there is any) using a forward-filling approach.
(C1-C2) The defined trigger-events in both channels along with the targets. 
Trigger-events were defined using our simulation methodology with IKF, dwell time of 100~ms, and buffer-period of 1000~ms.
The fourth row contains five subplots, each representing an individual target from the signal above with the respective defined trigger-event. 
Angular offsets are annotated above each subplot.}

\label{fig:1018}

\end{figure*}

First, occasional failures in eye-tracker performance led to Not-a-number(NaN) values in gaze data. 
Refer to Supplementary Fig. \ref{fig:1018} for an example of this scenario.
Plot (A1) and (A2) represent horizontal and vertical channels of an eye-tracking signal from GazeBase RAN data respectively.
There are five target points that the participants were supposed to follow. 
This portion of the recording contains NaNs.
To simulate our methodology, we handled the NaNs before applying any classification algorithms, as shown in plots (B1) and (B2).
Our method of handling these NaNs by referencing earlier values inadvertently introduced larger errors. 
We can see in the third target (light-yellow plot), that in the raw condition maximum portion of the gaze data is missing.
After handling NaNs in real-time, we define a trigger-event that has an angular offset of 19.82 dva.

\begin{figure*}[htbp]
\centering
\includegraphics[width=\textwidth]{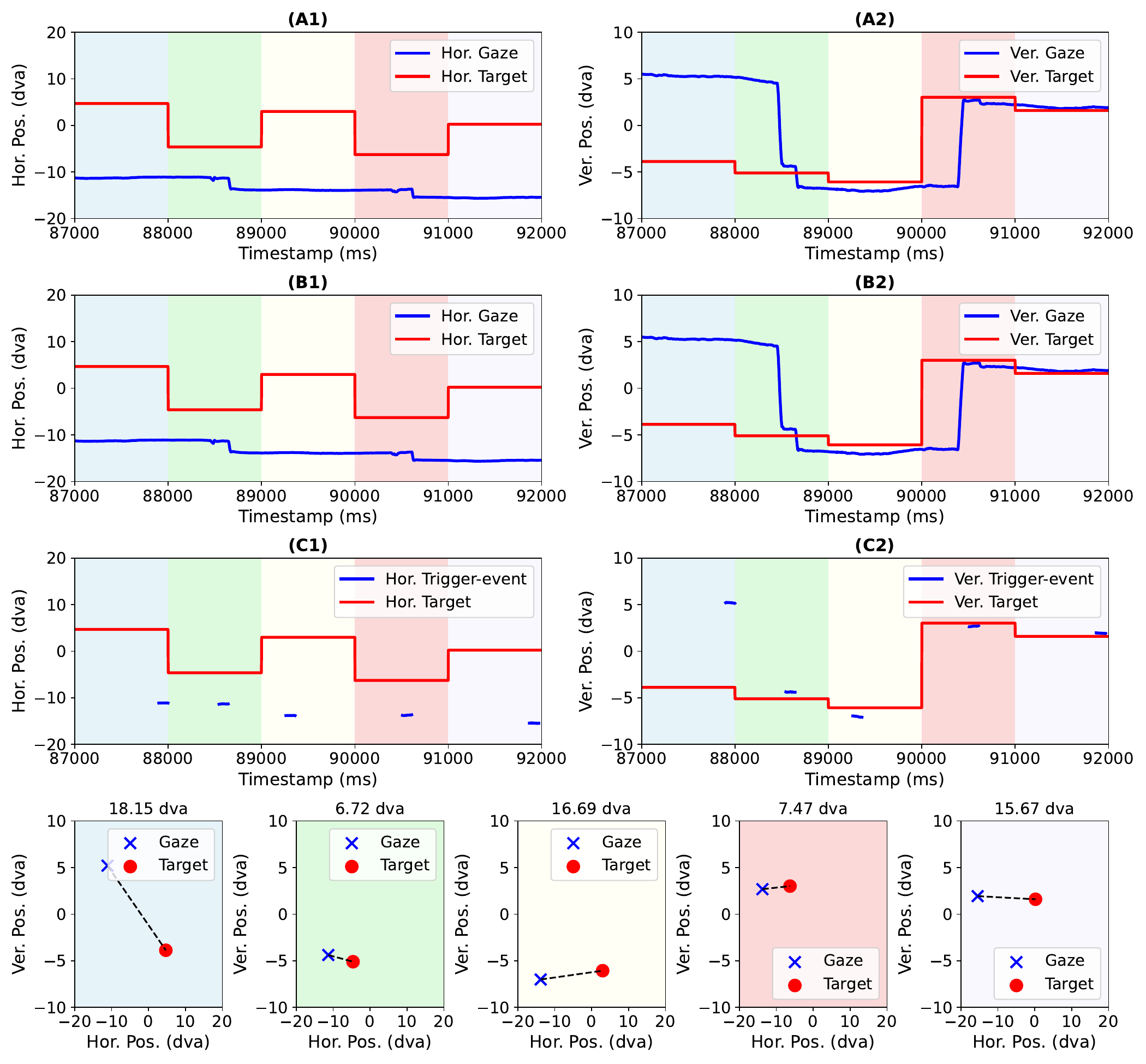}
\caption{Supplementary Fig. 2. Details are same as Figure \ref{fig:1018}}
\label{fig:1044}

\end{figure*}

Second, a consistent offset is observed between the gaze and target positions from the beginning of the recording. 
Supplementary Fig. \ref{fig:1044} illustrates an instance of this scenario.
Notably, the horizontal channel consistently exhibits an additional offset relative to the targets, as depicted in plot (A1).
Because of this situation, the trigger-event from the first target (light-blue plot) yields a high angular offset of 18.15 dva, high angular offset continues to the rest of the targets as well.
This finding suggests a potential issue during the data collection phase for this specific participant, which could be a calibration error.
This scenario is common in several recordings. 

\section{Success Rate of the Simulation Methodology under Varying Constraints}

Supplementary Table \ref{tab:acc} presents an extended version of the results from Table 1 in the main article, detailing the comparison of success rates (\%) between algorithms across different dwell times and buffer-periods. 

\begin{table}[ht!]
\centering
\caption{Comparison of success rate(\%) between algorithms for various dwell time and buffer-period. Higher values indicate better performance.}
\label{tab:acc}
\begin{tabular}{ccccc}
\hline
\multirow{2}{*}{Buffer-period (ms)} & \multirow{2}{*}{Dwell time (ms)} & \multicolumn{3}{c}{Algorithms} \\ \cline{3-5}
    &   &  IVT  &  IDT  &  IKF  \\ \hline
    
\multirow{5}{*}{1000} 
& 100 & 87.8  & 99.4  & 97.2    \\ \cline{2-5} 
& 150 & 83.4  & 97.2  & 94.1    \\ \cline{2-5} 
& 200 & 78.2  & 90.2  & 89.6    \\ \cline{2-5} 
& 250 & 71.1  & 73.2  & 82.5    \\ \cline{2-5} 
& 300 & 64.2  & 56.3  & 75.3    \\ 
\hline

\multirow{5}{*}{900} 
& 100 & 87.5  & 99.3  & 97.0    \\ \cline{2-5} 
& 150 & 82.4  & 96.5  & 93.7    \\ \cline{2-5} 
& 200 & 76.0  & 86.8  & 87.9    \\ \cline{2-5} 
& 250 & 66.6  & 63.1  & 78.4    \\ \cline{2-5} 
& 300 & 57.1  & 43.3  & 68.7    \\ 
\hline

\multirow{5}{*}{800} 
& 100 & 86.9 & 99.1 & 96.8 \\ \cline{2-5} 
& 150 & 81.3 & 95.4 & 93.0 \\ \cline{2-5} 
& 200 & 73.5 & 81.1 & 85.5 \\ \cline{2-5} 
& 250 & 60.2 & 48.4 & 72.3 \\ \cline{2-5} 
& 300 & 47.3 & 27.6 & 58.5 \\ 
\hline

\multirow{5}{*}{700} 
& 100 & 85.8 & 98.7 & 96.3 \\ \cline{2-5} 
& 150 & 79.6 & 93.5 & 91.9 \\ \cline{2-5} 
& 200 & 69.5 & 71.6 & 81.5 \\ \cline{2-5} 
& 250 & 45.3 & 29.1 & 56.7 \\ \cline{2-5} 
& 300 & 22.4 & 11.5 & 30.0\\ 
\hline

\multirow{5}{*}{600} 
& 100 & 84.5 & 97.9 & 95.6 \\ \cline{2-5} 
& 150 & 77.3 & 89.8 & 89.4 \\ \cline{2-5} 
& 200 & 56.3 & 58.1 & 66.1 \\ \cline{2-5} 
& 250 & 19.9 & 14.5 & 24.3 \\ \cline{2-5} 
& 300 & 11.1 & 5.1 & 14.2 \\ 
\hline

\multirow{5}{*}{500} 
& 100 & 82.6 & 96.5 & 94.2 \\ \cline{2-5} 
& 150 & 73.0 & 86.4 & 84.3 \\ \cline{2-5} 
& 200 & 16.0 & 50.4 & 51.0\\ \cline{2-5} 
& 250 & 10.4 & 7.8 & 12.4\\ \cline{2-5} 
& 300 & 3.8 & 3.4 & 3.6 \\ 
\hline

\multirow{5}{*}{400} 
& 100 & 80.7 & 94.8 & 91.7 \\ \cline{2-5} 
& 150 & 71.0 & 84.2 & 80.1 \\ \cline{2-5} 
& 200 & 39.0 & 46.2 & 40.2 \\ \cline{2-5} 
& 250 & 7.8 & 7.4 & 7.6 \\ \cline{2-5} 
& 300 & 3.7 & 3.3 & 3.6\\ 
\hline

\end{tabular}
\end{table}

\bibliographystyle{unsrt}
\bibliography{reference.bib}